\documentclass[twocolumn, aps, pra, 10pt]{revtex4-2}

\usepackage{graphicx}

\usepackage{amsmath, amssymb, amsfonts}

\usepackage{hyperref}
\usepackage{color}

\begin{document}

    \title{Synchronization of a superconducting qubit to an optical field mediated by a mechanical resonator}
    \author{Roson Nongthombam}
        \email{n.roson@iitg.ac.in}
        \affiliation{Department of Physics, Indian Institute of Technology Guwahati, Guwahati-781039, India}
    \author{Sampreet Kalita}
        \email{sampreet@iitg.ac.in}
        \affiliation{Department of Physics, Indian Institute of Technology Guwahati, Guwahati-781039, India}
    \author{Amarendra K. Sarma}
        \email{aksarma@iitg.ac.in}
        \affiliation{Department of Physics, Indian Institute of Technology Guwahati, Guwahati-781039, India}

    \date{\today}

    \begin{abstract}
        We study the synchronization of a superconducting qubit to an external optical field via a mechanical resonator in a hybrid optoelectromechanical system.
        The quantum trajectory method is employed to investigate synchronization.
        The bistability in one of the qubit polarization vectors, where the qubit rotates about the polarization vector, is observed for a single quantum trajectory run.
        The rotation in one of the stable states is synced with the external optical drive.
        When the number of trajectories is significantly increased, the qubit no longer displays bistability.
        However, synchronization with less quantum fluctuations is still observed.
        The scheme could be used to transfer the phase of the microwave qubit's rotation to a long-lived optical photon through synchronization, which may find applications in long-distance quantum communication.
        Also, this hybrid system can be used to study quantum synchronization.
    \end{abstract}

    \maketitle


    \section{Introduction}
        \label{sec:intro}
        Exchange of information between a microwave signal and an optical one is at the heart of realizing long-distance quantum communication.
        To this end, individual opto- and electro-mechanical systems that can effectively store and retrieve the optical and electrical information in the mechanical degrees of freedom have gathered a good amount of interest in the past decades \cite{JPhysConfSer.264.012025, RevModPhys.86.1391, NatNanotechnol.13.11, RevModPhys.93.025005}.
        A wide range of experiments have also demonstrated the fabrication of mechanical oscillators coupled to optical cavities \cite{Nature.452.72, Nature.460.724, NatPhys.5.485, Nature.478.89, Nature.482.63}, and mechanical resonators integrated into microwave circuits \cite{RevSciInstrum.76.061101, NatPhys.4.555, Nature.471.204, Nature.475.359}.
        This has led to hybrid optoelectromechanical (OEM) systems, consisting of both optical and electrical elements coupled via a common mechanical membrane, which facilitate the study of quantum properties like entanglement \cite{PhysRevA.84.042342, PhysRevLett.109.130503}, state-transfer \cite{PhysRevA.92.043845, NatPhys.10.321}, and ground-state cooling \cite{PhysRevA.103.033516, PhysRevA.104.023509}.
        Such hybrid OEM systems can also be integrated into solid-state devices that can communicate to similar setups via optical photons.
        However, a superconducting element such as a qubit cannot be directly coupled to an optical element as it would lead to the destruction of the superconductivity.
        To overcome this limitation, a mechanical oscillator inside a electromechanical circuit, seamlessly coupled to a superconducting qubit at one end and to an optical cavity on the other, can be used as a mediator of information exchange between the qubit and the optical photons \cite{RevModPhys.93.025005}.
        Such a form of coupling has been explored both theoretically \cite{PhysRevB.69.125339, PhysRevA.76.042336, NewJPhys.10.085018, NewJPhys.10.085019,PhysRevResearch.2.023335} and experimentally \cite{Nature.459.960, Nature.494.211, Science.358.199}.
        A bidirectional transfer of information between a qubit and a mechanical oscillator via a piezoelectric coupling and transferring the same to an optical field via an optomechanical coupling is demonstrated in \cite{Nature.588.599}.

        An advantage of encoding information in optical photons is that they can retain their quantumness even at room temperatures, which facilitates long-distance quantum state transfer.
        Naturally, encoding the qubit state in an optical photon can pave the path towards building efficient quantum networks.
        A possible way to achieve this encoding is by correlating the rotation of a qubit state to a controllable and readable optical element via synchronization \textemdash{} a natural tendency of oscillators to adjust their rhythms sympathetically \cite{PBL.Sync.Strogatz}.
        This classical phenomenon \cite{SciRep.5.11548} has gathered a good amount of interest in a variety of quantum systems \cite{PhysRevLett.111.234101, PhysRevA.91.061401, AnnPhysBerl.527.131, PhysRevLett.117.073601, CommunNonlinearSciNumerSimulat.42.121, PhysRevA.99.043804, PhysRevResearch.2.023026}.
        The synchronization of a superconducting qubit with a driven microwave cavity \cite{PhysRevLett.100.014101}, and a mechanical oscillator to an external drive in an optomechanical system \cite{PhysRevA.95.053858} has also been explored.
        In this work, we investigate synchronization in a hybrid system consisting of a microwave circuit with the superconducting qubit coupled mechanically to a driven optomechanical cavity.
        For a single quantum trajectory run, we observe bistability in one of the qubit polarization vectors, where the qubit rotates about the polarization vector.
        Through the qubit-mechanical and mechanical-optical cavity coupling, the phase of the qubit rotation in one of the stable states synchronizes with the phase of the reference laser.
        We further analyze the dynamics of our system for a larger number of quantum trajectories and observe a similar pattern in the synchronization but with a lesser amount of fluctuations.
        However, in this case, we no longer observe bistability in the qubit state.
        This work presents an alternative approach to transduce information between a microwave and an optical signal by synchronizing a superconducting qubit to an external optical field using a hybrid OEM system.
        
        The paper is organized as follows. We describe the hybrid system under study in Sec. \ref{sec:system}. 
        The bistability and rotation of the qubit as well as the self-sustained oscillation of the mechanical oscillator is studied numerically in Sec. \ref{sec:bis_rot_osc}.
        In Sec. \ref{sec:sync}, we show the synchronization of the qubit to the external drive obtained by simulating both single trajectories and averaging over a large number of trajectories.
        Finally, we conclude by summarizing our work in Sec. \ref{sec:conc}.  


    \section{The Hybrid System}
        \label{sec:system}
\begin{figure}[t]
            \centering
            \includegraphics[width=0.48\textwidth]{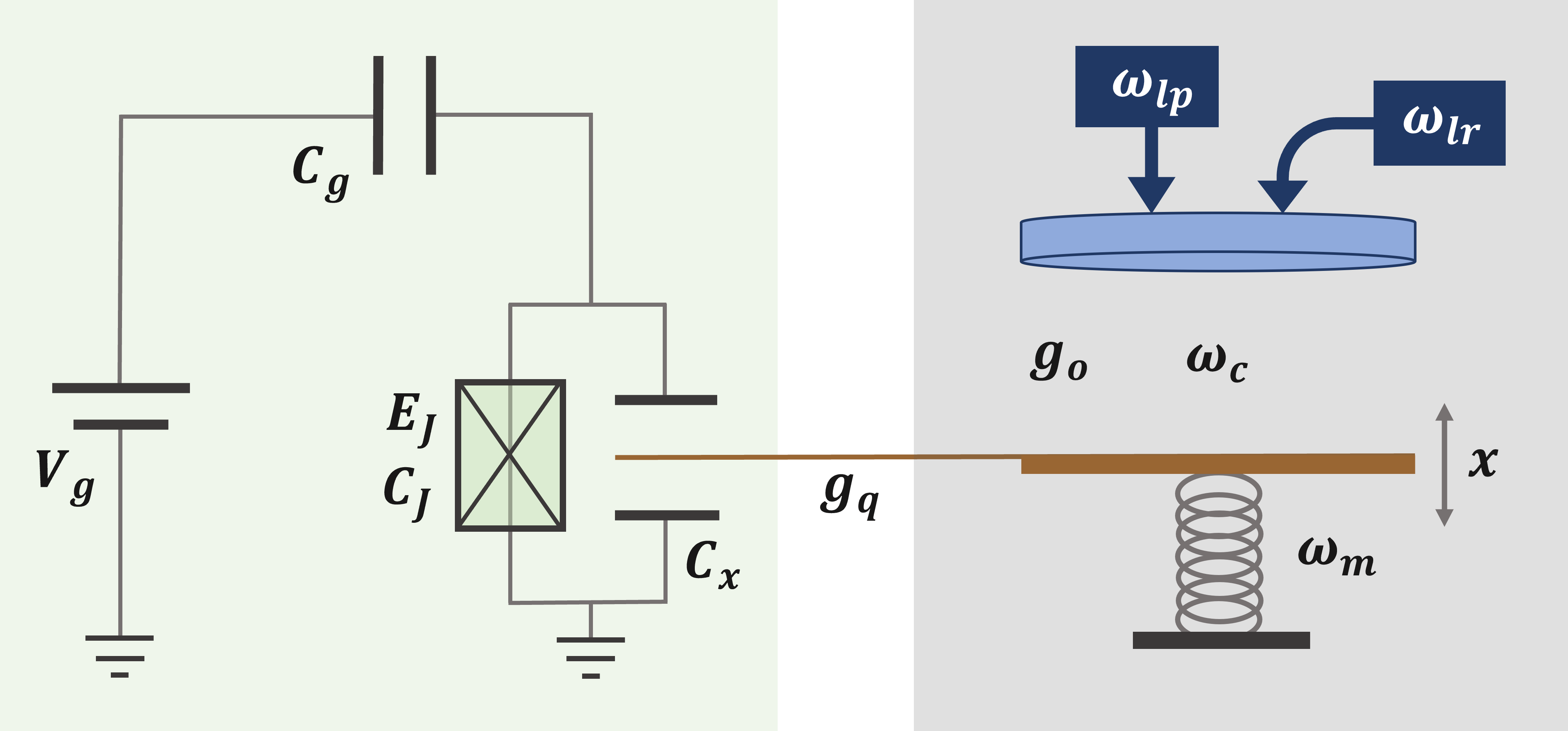}
            \caption{Schematics of a mechanically-mediated OEM system with a superconducting qubit (left) and an optomechanical cavity (right).
            The qubit is capacitively coupled to a mechanical oscillator, which acts as an end-mirror of an optical cavity.}
            \label{fig:system}
        \end{figure}

        We consider a hybrid OEM system consisting of a superconducting qubit coupled to a mechanical oscillator, which in turn is parametrically coupled to an optical cavity as depicted in Fig. \ref{fig:system}.
        The Hamiltonian of our system can therefore be expressed in three distinctive parts \textemdash{} (i) $H_{qm}$, consisting of the electromechanical dynamics of the qubit and its interaction with the mechanical oscillator, (ii) $H_{om}$, for the driven optical cavity and the optomechanical interaction, and (iii) $\hat{H}_{m}$, the energy of the self-sustained mechanical motion.
        In the rotating frame of the laser frequency, the individual Hamiltonians (in units of $\hbar$) read as (refer Appendix \ref{app:ham_qubit})
        \begin{subequations}
            \begin{eqnarray}
                \label{eqn:ham_qm}
                \hat{H}_{qm} &=& -\frac{E_{J}}{2} \sigma_{x} + g_{q} \left( \hat{b}^{\dagger} + \hat{b} \right) \sigma_{z}, \\ 
                 \label{eqn:ham_m}
                \hat{H}_{m} &=& \omega_{m} \hat{b}^{\dagger} \hat{b}, \\
                \label{eqn:ham_om}
                \hat{H}_{om} &=& - \Delta \hat{a}^{\dagger} \hat{a} - g_{o} \hat{a}^{\dagger} \hat{a} \left( \hat{b}^{\dagger} + \hat{b} \right) + iA_{lp} \left( \hat{a}^{\dagger} - \hat{a} \right).
             \end{eqnarray}
        \end{subequations}
        
        Here, $\hat{a}$ ($\hat{a}^{\dagger}$) and $\hat{b}$ ($\hat{b}^{\dagger}$) are the annihilation (creation) operators of the optical photons inside the cavity and the phonons of the mechanical oscillator respectively.
        The first term of Eq. \eqref{eqn:ham_qm} denotes the qubit state with Josephson energy $E_{J}$; the second term represents the qubit-phonon interaction with a coupling strength of $g_{q}$.
        The term in Eq. \eqref{eqn:ham_m} indicates the energy of mechanical mode oscillating with frequency, $\omega_m$. 
        The first and the final terms of Eq. \eqref{eqn:ham_om} denote the energies of the optical cavity and the driving laser, where $\Delta = \omega_{lp} - \omega_{c}$ is the detuning of the laser frequency $\omega_{lp}$ from the cavity resonance frequency $\omega_{c}$, and $A_{lp}$ is the laser amplitude.
        The central term denotes the optomechanical interaction of strength $g_{o}$.
        In addition to the primary laser drive ($A_{lp}$, $\omega_{lp}$), we add a reference drive ($A_{lr}$, $\omega_{lr}$) with which we sync the qubit's rotational state.
        The corresponding Hamiltonian (in units of $\hbar$) in the rotating frame of the primary laser drive is given by 
        \begin{equation}
            \label{eqn:ham_lr}
            \hat{H}_{lr} = i A_{lr} \left( \hat{a}^{\dagger} e^{-i \Omega t} - \hat{a} e^{i \Omega t} \right),
        \end{equation} 
        where $\Omega = \omega_{lr} - \omega_{lp}$ is the detuning of the reference laser.
        The total Hamiltonian of the system therefore becomes
        \begin{equation}
            \label{eqn:ham}
            \hat{H} = \hat{H}_{qm} + \hat{H}_{om} + \hat{H}_{m} + \hat{H}_{lr}.
        \end{equation}
        
        The complete dynamics of the system consists of the dissipation of the mechanical oscillator as well as the cavity resonator and can be described by the Lindblad master equation \cite{TaylorFrancis.QuantumOptomechanics.Bowen},
        \begin{equation}
            \label{eqn:master}
            \dot{\hat{\rho}} = -i \left[ \hat{H},\hat{\rho} \right] + \kappa \mathcal{L} \left[ \hat{a} \right] + \gamma \mathcal{L} \left[ \hat{b} \right],
        \end{equation}
        where $\mathcal{L} [ \hat{o} ] = ( 2 \hat{o} \hat{\rho} \hat{o}^{\dagger} - \hat{o}^{\dagger} \hat{o} \hat{\rho} - \hat{\rho} \hat{o}^{\dagger} \hat{o} ) / 2$ with $\hat{o} \in \{ \hat{a}, \hat{b} \}$.
        Here, while writing Eq. \ref{eqn:master}, we assume (i) the thermal phonons and photons are negligible \cite{PhysRevA.95.053858}, and (ii) the qubit decays much slower than the mechanical and cavity resonators \cite{PhysRevLett.100.014101}.

    \section{Bistability, Rotation and Self-sustained Oscillation}
        \label{sec:bis_rot_osc}  
        \begin{figure}[ht]
            \centering
            \includegraphics[width=0.48\textwidth]{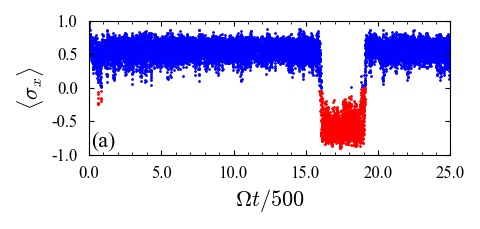}
            \includegraphics[width=0.48\textwidth]{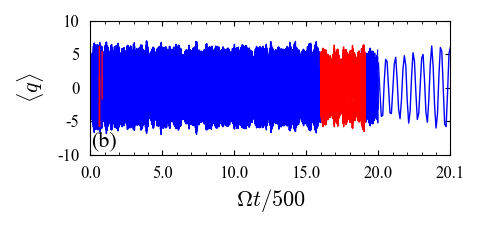}
            \includegraphics[width=0.27\textwidth]{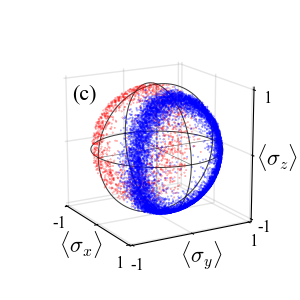}
            \includegraphics[width=0.19\textwidth]{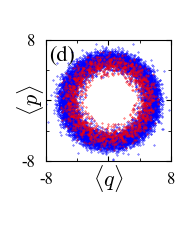}
            \caption{(Color online) (a) Bistability of the qubit along the polarization vector $\langle \sigma_{x} \rangle$.
            (b) Fluctuating amplitude oscillation of the mechanical resonator.
            (c) Bloch sphere representation of the qubit rotation in the two stable states (blue and red states).
            (d) Overlapping limit cycles of the mechanical oscillator corresponding to the two bistable states of the qubit.
            The parameters used are $( E_{J}, g_{q}, \Delta, g_{o}, A_{lp}, A_{lr}, \Omega, \kappa, \gamma ) = ( 1.2, 0.04, 1.0, 0.38, 0.6, 0.08, 1.0, 1.4, 0.015 ) \times \omega_{m}$.}
            \label{fig:bis_rot_osc}
        \end{figure}

        \begin{figure}[ht]
            \centering
            \includegraphics[width=0.19\textwidth]{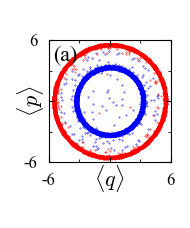}
            \includegraphics[width=0.27\textwidth]{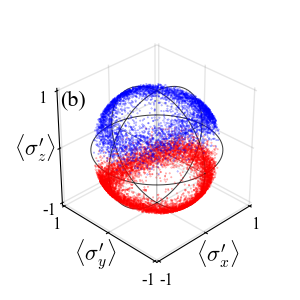}
            \caption{(Color online) (a) Two well-separated and distinct limit cycles of the mechanical oscillator when it is directly driven.
            In the absence of the optical cavity, the mechanical-qubit system shows less fluctuation.
            (b) Rotation in the qubit basis.
            The parameters used are same as those in Fig. \ref{fig:bis_rot_osc}.}
            \label{fig:bis_rot_osc_qubit}
        \end{figure}
        We simulate the dynamics of the hybrid system using the quantum trajectory method \cite{AmJPhys.70.719} numerically \cite{ComputPhysCommun.183.1760, ComputPhysCommun.184.1234}.
        The results are obtained at irrational moments of the optical drive phase $\Omega t$ and for a single quantum trajectory.
        A single quantum trajectory evolution is equivalent to one experimental run.
        
        For the qubit, we observe a bistable behaviour \textemdash{} two metastable states of its polarization vector $\langle \sigma_{x} \rangle$ \textemdash{} as shown in Fig. \ref{fig:bis_rot_osc} (a).
        The two other polarization vectors $\langle \sigma_{y} \rangle$ and $\langle \sigma_{z} \rangle$ oscillate about the origin.
        When the qubit is in either of the metastable states, its Bloch vector precesses about the x-axis, as can be seen in Fig. \ref{fig:bis_rot_osc} (c).
        The mechanical oscillator also undergoes self-sustained oscillations due to the blue-detuned laser drive $( \Delta = \omega_{m} )$ of the optomechanical cavity as shown in Fig. \ref{fig:bis_rot_osc} (b). 
        Similar to the case of qubit bistability, we observe two different phase oscillations in the mechanical oscillator.
        It can be seen from Fig. \ref{fig:bis_rot_osc} (d) that the limit cycles corresponding to the mechanical oscillations follow the qubit bistability through the qubit-mechanical coupling and are very close to each other.
        Later, we find that the qubit rotation phase is synced to the self-sustained oscillations of the mechanics, which in turn is fixed by the frequency of the external reference field (refer Fig. \ref{fig:sync_phase}).
        
        In Fig. \ref{fig:bis_rot_osc}, we also observe broad oscillations in the qubit and the mechanical dynamics.
        The occurrence of this broadness can be attributed to two reasons.
        Firstly, the fluctuating radiation pressure force imparted by the external laser via the cavity photons induces fluctuations in the self-sustained motion of the mechanical oscillator.
        This results in the broadness of the phase-space limit cycles, which in turn, invokes the spreading of the qubit's metastable states via the qubit-mechanical coupling.
        When the mechanical oscillator is independently driven without the presence of an optical cavity, depicted by the Hamiltonian $\hat{H} = \hat{H}_{qm} + \hat{H}_{m}$, we find two well-separated and distinct limit cycles in the mechanical phase space as shown in Fig. \ref{fig:bis_rot_osc_qubit} (a).
        The metastable states of the qubit polarization vector $\langle \sigma_{x} \rangle$ are also thinner \cite{PhysRevLett.100.014101}.
        Nevertheless, some quantum noises still remain.
        Secondly, we have used a single trajectory run to obtain our results in Fig. \ref{fig:bis_rot_osc}.
        As we increase the number of trajectories, we get less scattered and thinner limit cycles \cite{PhysRevA.95.053858}.
        The fluctuations (thickness) of the synchronization phase plot is also reduced (refer Fig. \ref{fig:sync_ntraj}).
        This is for the fact that considering a larger number of trajectories is equivalent to taking the ensemble average of the trajectories which significantly cancels out the fluctuating noises.
        The effect of a large number of quantum trajectories in our hybrid system is also discussed in Sec. \ref{sec:sync}.
        In Fig. \ref{fig:bis_rot_osc_qubit} (b), we plot the Bloch vector in the frame of reference of the qubit, $\hat{H}_{qm}^{\prime} = \frac{E_{J}}{2} \sigma_{z}^{\prime} + g_{q} ( \hat{b}^{\dagger} + \hat{b} ) \sigma_{x}^{\prime}$ (refer Appendix \ref{app:ham_qubit}).
        Here too, we observe bistability along $\langle \sigma_{z}^{\prime} \rangle$ and rotation of the qubit about the z-axis.

        It is also worth mentioning that even though we are bound by the fluctuating radiation force of the optical photons, we are able to observe quantum synchronization of the qubit with the external optical field within this bound quantum noise.
        In what follows, we discuss this synchronous behaviour of the qubit rotation phase with the phase of the mechanical oscillation, which in turn, is in sync with that of the reference laser field.


    \section{Synchronization with the External Drive}
        \label{sec:sync}\begin{figure}[t]
            \centering
            \includegraphics[width=0.23\textwidth]{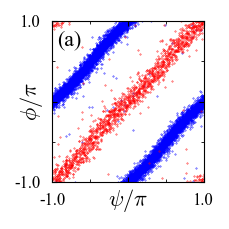}
            \includegraphics[width=0.23\textwidth]{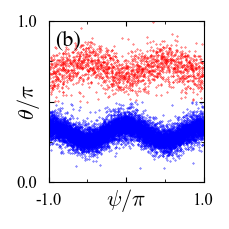}
            \includegraphics[width=0.23\textwidth]{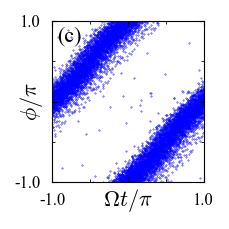}
            \includegraphics[width=0.23\textwidth]{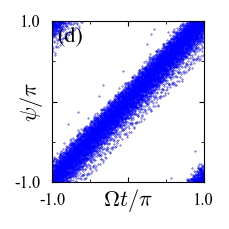}
            \includegraphics[width=0.23\textwidth]{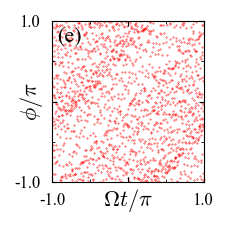}
            \includegraphics[width=0.23\textwidth]{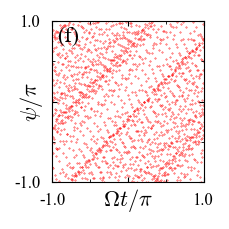}
            \caption{(Color online) Synchronization of the hybrid system.
            (a) Qubit phase ($\phi$) versus mechanical phase ($\psi$) plot.
            (b) Qubit phase ($\theta$) versus mechanical phase ($\psi$) plot.
            Phase synchronization of the two stable qubit states with the two limit cycles of the mechanical oscillator indicated by red and blue is clearly seen in (a) and (b).
            Synchronization of the qubit blue state with the external optical drive is shown in (c).
            Similarly, phase plot of the mechanical oscillator in the blue limit cycle and the external drive is plotted in (d).
            (e) Unsynchronization phase plot of the qubit red state versus the external drive.
            (d) Phase of the red limit cycle versus phase of the external drive phase.
            The parameters used are same as those in Fig. \ref{fig:bis_rot_osc}.}
            \label{fig:sync_phase}
        \end{figure}
        
        \begin{figure}[t]
            \centering
            \includegraphics[width=0.23\textwidth]{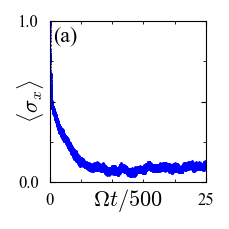}
            \includegraphics[width=0.23\textwidth]{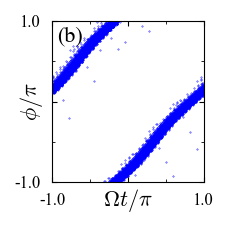}
            \includegraphics[width=0.23\textwidth]{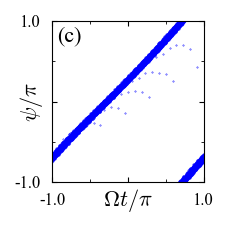}
            \includegraphics[width=0.23\textwidth]{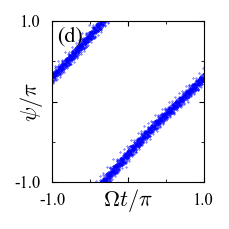}
            \caption{(Color online) Results for a large number of trajectories run.
            (a) Time evolution of qubit vector $\langle \sigma_{x} \rangle$. We no longer observe bistability. The qubit vector decays to a near zero state.
            (b) and (c) shows the less fluctuating synchronization of the qubit and the mechanical oscillator with the reference field, respectively.
            The parameters used are same as those in Fig. \ref{fig:bis_rot_osc}.
            (d) Phase plot of the mechanical oscillator and the drive field in a pure optomechanical system \cite{PhysRevA.95.053858}.
            In this case also we observe less fluctuations due to the large number of trajectories.}  
            \label{fig:sync_ntraj}
        \end{figure}

        Using the phase relations of the qubit, $\phi = \tan^{-1} ( \langle \sigma_{y} \rangle / \langle \sigma_{z} \rangle )$, $\theta = \tan^{-1} ( \langle \sigma_{y} \rangle / \{ \langle \sigma_{x} \rangle \sin{( \phi )} \} )$, and that of the mechanical oscillator, $\psi = \tan^{-1} ( \langle p \rangle / \langle q \rangle )$, we show their synchronization in Figs. \ref{fig:sync_phase} (a)-(b).
        It can be seen that the rotation of the $\langle \sigma_{x} \rangle > 0$ stable state (blue state) and the rotation of the $\langle \sigma_{x} \rangle < 0$ stable state (red state) are synced to the corresponding red and blue limit cycles of the mechanical oscillator.
        Also, we plot the variation in the phase of the qubit rotation and that of the mechanical oscillator with the phase of the reference laser field (mod $2 \pi$) in Figs. \ref{fig:sync_phase} (c)-(f).
        Here, we observe that the reference drive is synced only to one of the phases of the mechanical oscillator, and subsequently to one of the phases of the qubit rotation.
        We explain this as follows.
        The qubit and the mechanical oscillator have two different phases owing to bistability (red and blue states).
        However, the reference drive has only a single phase preference and therefore, displays monostable synchronization.
        For the same parameters as Fig. \ref{fig:bis_rot_osc}, we observe that the optical drive is synced to the blue stable state of the qubit rotation (Fig. \ref{fig:sync_phase} (c)) and blue limit cycle of the mechanical oscillator (Fig. \ref{fig:sync_phase} (d)).
        This synchronization is found to be more prominent, i.e., having less quantum noise, when the qubit spends more time in the blue state.
        For every single quantum trajectory run, we get different time span of the bistable states due to the inherent stochastic behaviour of quantum trajectories, and when the time span of blue state is more than the red one, we get a clear synchronization.
        In other words, the measure of fluctuations in synchronization serves as a means of knowing the time span of the stable states.
        The corresponding scenario when the drive is not synchronized (red state) is shown in Figs. \ref{fig:sync_phase} (e)-(f).
        A similar behaviour of synchronization is observed in the qubit basis. 
        Here, the phase relations change to $\phi = \tan^{-1}(\langle \sigma_{y}^{\prime} \rangle/\langle \sigma_{x}^{\prime} \rangle )$ and $\theta = tan^{-1} ( \langle \sigma_{y}^{\prime} \rangle / \{ \langle \sigma_{z}^{\prime} \rangle \sin{( \phi )} \} )$.

        In Fig. \ref{fig:sync_ntraj}, we show our results for a large number of quantum trajectories.
        The qubit polarization state $\langle \sigma_{x} \rangle$ averages out to a near-zero state and the quantum noise in the synchronization plot is also significantly reduced.
        However, the amplitude of the qubit rotation decreases and the limit cycle of the mechanical oscillator shrinks towards the origin as the number of trajectories increases.
        It is also worth mentioning here that although the actual dynamics of the system can follow a very complicated periodic structure, yet at specific intervals of time, simple patterns emerge as can be seen from Fig. \ref{fig:bis_rot_osc} and Fig. \ref{fig:sync_phase}.
        Our results also indicate that at irrational moments of $\Omega t$, a limit-cycle dynamics is observed leading to synchronization between the qubit rotation and the reference drive.
        Further, for single trajectory runs, our system could provide a novel platform to explore the synchronization dynamics of individual quantum systems for applications in communication via transduction.

    
    \section{Conclusion}
        \label{sec:conc}
        In conclusion, our work has numerically shown the synchronization of a superconducting qubit to an external optical field via a mechanical resonator.
        We observed bistability in the qubit for a single quantum trajectory run, and the rotation in the $\langle \sigma_{x} \rangle > 0$ stable state is synced to the external optical drive.
        When the number of trajectories is significantly increased, we find that the qubit no longer displays bistability.
        However, synchronization with less quantum fluctuations is still observed.
        Our scheme paves the way to synchronize the phase of rotation for one of the stable states of the qubit to the reference drive.
        It facilitates the transduction of information between the qubit and the optical system which may find applications in long-distance quantum communication.
        Moreover, our scheme could be used to explore synchronization in the quantum regime.


    \section*{Acknowledgement}
        RN gratefully acknowledges support of a research fellowship from CSIR, Govt. of India.
        SK would like to acknowledge MHRD, Government of India for providing financial support for his research via the PMRF scheme.
        Simulations are done using the QuTiP library \cite{ComputPhysCommun.184.1234}.


    \appendix

    \section{Qubit Hamiltonian}
        \label{app:ham_qubit}
        The Hamiltonian of a superconducting qubit coupled to a mechanical oscillator is given by \cite{PhysRevA.104.023509}
        \begin{align} 
            \label{eqn:A1}
            \hat{H}_{qm} = -\frac{\epsilon}{2} \sigma_{z} - \frac{E_{J}}{2} \sigma_{x} + g_{q} \left( \hat{b}^{\dagger} + \hat{b} \right) \sigma_{z},
        \end{align}
        where $\epsilon = 8 E_{c} \Delta N$, $\Delta N$ is the gate charge number near the sweet point ($\Delta N = 0$) and $E_{c}$ is the charging energy.
        By substituting $\Delta N = 0$ in Eq. \eqref{eqn:A1}, we get Eq. \ref{eqn:ham_qm}.
        Going to the basis state of the qubit, we obtain
        \begin{eqnarray}
            \label{eqn:A2}
            \hat{H}_{qm} &=& \frac{\sqrt{\epsilon^{2} + E_{J}^{2}}}{2}\sigma_{z}^{\prime} \nonumber \\
            && + g_{q} \left( \hat{b}^{\dagger} + \hat{b} \right) \left( \sigma_{x}^{\prime} \cos{\phi} - \sigma_{z}^{\prime} \sin{\phi} \right),
        \end{eqnarray}
        where $\tan{\phi} = \epsilon / E_{J}$.
        At the sweet point $\epsilon = 0$, i.e., $\sin{\phi} = 0$ and $\cos{\phi} = 1$.
        Therefore, 
        \begin{equation}
            \label{eqn:A3}
            \hat{H}_{qm} = \frac{E_{J}}{2}\sigma_{z}^{\prime} + g_{q} \left( \hat{b}^{\dagger} + \hat{b} \right) \sigma_{x}^{\prime}.
        \end{equation}

    \bibliography{references}

\begin{thebibliography}{43}%
\makeatletter
\providecommand \@ifxundefined [1]{%
 \@ifx{#1\undefined}
}%
\providecommand \@ifnum [1]{%
 \ifnum #1\expandafter \@firstoftwo
 \else \expandafter \@secondoftwo
 \fi
}%
\providecommand \@ifx [1]{%
 \ifx #1\expandafter \@firstoftwo
 \else \expandafter \@secondoftwo
 \fi
}%
\providecommand \natexlab [1]{#1}%
\providecommand \enquote  [1]{``#1''}%
\providecommand \bibnamefont  [1]{#1}%
\providecommand \bibfnamefont [1]{#1}%
\providecommand \citenamefont [1]{#1}%
\providecommand \href@noop [0]{\@secondoftwo}%
\providecommand \href [0]{\begingroup \@sanitize@url \@href}%
\providecommand \@href[1]{\@@startlink{#1}\@@href}%
\providecommand \@@href[1]{\endgroup#1\@@endlink}%
\providecommand \@sanitize@url [0]{\catcode `\\12\catcode `\$12\catcode
  `\&12\catcode `\#12\catcode `\^12\catcode `\_12\catcode `\%12\relax}%
\providecommand \@@startlink[1]{}%
\providecommand \@@endlink[0]{}%
\providecommand \url  [0]{\begingroup\@sanitize@url \@url }%
\providecommand \@url [1]{\endgroup\@href {#1}{\urlprefix }}%
\providecommand \urlprefix  [0]{URL }%
\providecommand \Eprint [0]{\href }%
\providecommand \doibase [0]{https://doi.org/}%
\providecommand \selectlanguage [0]{\@gobble}%
\providecommand \bibinfo  [0]{\@secondoftwo}%
\providecommand \bibfield  [0]{\@secondoftwo}%
\providecommand \translation [1]{[#1]}%
\providecommand \BibitemOpen [0]{}%
\providecommand \bibitemStop [0]{}%
\providecommand \bibitemNoStop [0]{.\EOS\space}%
\providecommand \EOS [0]{\spacefactor3000\relax}%
\providecommand \BibitemShut  [1]{\csname bibitem#1\endcsname}%
\let\auto@bib@innerbib\@empty
\bibitem [{\citenamefont {Regal}\ and\ \citenamefont
  {Lehnert}(2011)}]{JPhysConfSer.264.012025}%
  \BibitemOpen
  \bibfield  {author} {\bibinfo {author} {\bibfnamefont {C.~A.}\ \bibnamefont
  {Regal}}\ and\ \bibinfo {author} {\bibfnamefont {K.~W.}\ \bibnamefont
  {Lehnert}},\ }\href {https://doi.org/10.1088/1742-6596/264/1/012025}
  {\bibfield  {journal} {\bibinfo  {journal} {J. Phys.: Conf. Ser.}\ }\textbf
  {\bibinfo {volume} {264}},\ \bibinfo {pages} {012025} (\bibinfo {year}
  {2011})}\BibitemShut {NoStop}%
\bibitem [{\citenamefont {Aspelmeyer}\ \emph {et~al.}(2014)\citenamefont
  {Aspelmeyer}, \citenamefont {Kippenberg},\ and\ \citenamefont
  {Marquardt}}]{RevModPhys.86.1391}%
  \BibitemOpen
  \bibfield  {author} {\bibinfo {author} {\bibfnamefont {M.}~\bibnamefont
  {Aspelmeyer}}, \bibinfo {author} {\bibfnamefont {T.~J.}\ \bibnamefont
  {Kippenberg}},\ and\ \bibinfo {author} {\bibfnamefont {F.}~\bibnamefont
  {Marquardt}},\ }\href {https://doi.org/10.1103/RevModPhys.86.1391} {\bibfield
   {journal} {\bibinfo  {journal} {Rev. Mod. Phys.}\ }\textbf {\bibinfo
  {volume} {86}},\ \bibinfo {pages} {1391} (\bibinfo {year}
  {2014})}\BibitemShut {NoStop}%
\bibitem [{\citenamefont {Midolo}\ \emph {et~al.}(2018)\citenamefont {Midolo},
  \citenamefont {Schliesser},\ and\ \citenamefont
  {Fiore}}]{NatNanotechnol.13.11}%
  \BibitemOpen
  \bibfield  {author} {\bibinfo {author} {\bibfnamefont {L.}~\bibnamefont
  {Midolo}}, \bibinfo {author} {\bibfnamefont {A.}~\bibnamefont {Schliesser}},\
  and\ \bibinfo {author} {\bibfnamefont {A.}~\bibnamefont {Fiore}},\ }\href
  {https://doi.org/10.1038/s41565-017-0039-1} {\bibfield  {journal} {\bibinfo
  {journal} {Nat. Nanotechnol.}\ }\textbf {\bibinfo {volume} {13}},\ \bibinfo
  {pages} {11} (\bibinfo {year} {2018})}\BibitemShut {NoStop}%
\bibitem [{\citenamefont {Blais}\ \emph {et~al.}(2021)\citenamefont {Blais},
  \citenamefont {Grimsmo}, \citenamefont {Girvin},\ and\ \citenamefont
  {Wallraff}}]{RevModPhys.93.025005}%
  \BibitemOpen
  \bibfield  {author} {\bibinfo {author} {\bibfnamefont {A.}~\bibnamefont
  {Blais}}, \bibinfo {author} {\bibfnamefont {A.~L.}\ \bibnamefont {Grimsmo}},
  \bibinfo {author} {\bibfnamefont {S.~M.}\ \bibnamefont {Girvin}},\ and\
  \bibinfo {author} {\bibfnamefont {A.}~\bibnamefont {Wallraff}},\ }\href
  {https://doi.org/10.1103/RevModPhys.93.025005} {\bibfield  {journal}
  {\bibinfo  {journal} {Rev. Mod. Phys.}\ }\textbf {\bibinfo {volume} {93}},\
  \bibinfo {pages} {025005} (\bibinfo {year} {2021})}\BibitemShut {NoStop}%
\bibitem [{\citenamefont {Thompson}\ \emph {et~al.}(2008)\citenamefont
  {Thompson}, \citenamefont {Zwickl}, \citenamefont {Jayich}, \citenamefont
  {Marquardt}, \citenamefont {Girvin},\ and\ \citenamefont
  {Harris}}]{Nature.452.72}%
  \BibitemOpen
  \bibfield  {author} {\bibinfo {author} {\bibfnamefont {J.~D.}\ \bibnamefont
  {Thompson}}, \bibinfo {author} {\bibfnamefont {B.~M.}\ \bibnamefont
  {Zwickl}}, \bibinfo {author} {\bibfnamefont {A.~M.}\ \bibnamefont {Jayich}},
  \bibinfo {author} {\bibfnamefont {F.}~\bibnamefont {Marquardt}}, \bibinfo
  {author} {\bibfnamefont {S.~M.}\ \bibnamefont {Girvin}},\ and\ \bibinfo
  {author} {\bibfnamefont {J.~G.~E.}\ \bibnamefont {Harris}},\ }\href
  {https://doi.org/10.1038/nature06715} {\bibfield  {journal} {\bibinfo
  {journal} {Nature}\ }\textbf {\bibinfo {volume} {452}},\ \bibinfo {pages}
  {72} (\bibinfo {year} {2008})}\BibitemShut {NoStop}%
\bibitem [{\citenamefont {Gr\"{o}blacher}\ \emph
  {et~al.}(2009{\natexlab{a}})\citenamefont {Gr\"{o}blacher}, \citenamefont
  {Hammerer}, \citenamefont {Vanner},\ and\ \citenamefont
  {Aspelmeyer}}]{Nature.460.724}%
  \BibitemOpen
  \bibfield  {author} {\bibinfo {author} {\bibfnamefont {S.}~\bibnamefont
  {Gr\"{o}blacher}}, \bibinfo {author} {\bibfnamefont {K.}~\bibnamefont
  {Hammerer}}, \bibinfo {author} {\bibfnamefont {M.~R.}\ \bibnamefont
  {Vanner}},\ and\ \bibinfo {author} {\bibfnamefont {M.}~\bibnamefont
  {Aspelmeyer}},\ }\href {https://doi.org/10.1038/nature08171} {\bibfield
  {journal} {\bibinfo  {journal} {Nature}\ }\textbf {\bibinfo {volume} {460}},\
  \bibinfo {pages} {724} (\bibinfo {year} {2009}{\natexlab{a}})}\BibitemShut
  {NoStop}%
\bibitem [{\citenamefont {Gr\"{o}blacher}\ \emph
  {et~al.}(2009{\natexlab{b}})\citenamefont {Gr\"{o}blacher}, \citenamefont
  {Hertzberg}, \citenamefont {Vanner}, \citenamefont {Cole}, \citenamefont
  {Gigan}, \citenamefont {Schwab},\ and\ \citenamefont
  {Aspelmeyer}}]{NatPhys.5.485}%
  \BibitemOpen
  \bibfield  {author} {\bibinfo {author} {\bibfnamefont {S.}~\bibnamefont
  {Gr\"{o}blacher}}, \bibinfo {author} {\bibfnamefont {J.~B.}\ \bibnamefont
  {Hertzberg}}, \bibinfo {author} {\bibfnamefont {M.~R.}\ \bibnamefont
  {Vanner}}, \bibinfo {author} {\bibfnamefont {G.~D.}\ \bibnamefont {Cole}},
  \bibinfo {author} {\bibfnamefont {S.}~\bibnamefont {Gigan}}, \bibinfo
  {author} {\bibfnamefont {K.~C.}\ \bibnamefont {Schwab}},\ and\ \bibinfo
  {author} {\bibfnamefont {M.}~\bibnamefont {Aspelmeyer}},\ }\href
  {https://doi.org/10.1038/nphys1301} {\bibfield  {journal} {\bibinfo
  {journal} {Nat. Phys.}\ }\textbf {\bibinfo {volume} {5}},\ \bibinfo {pages}
  {485} (\bibinfo {year} {2009}{\natexlab{b}})}\BibitemShut {NoStop}%
\bibitem [{\citenamefont {Chan}\ \emph {et~al.}(2011)\citenamefont {Chan},
  \citenamefont {Alegre}, \citenamefont {Safavi-Naeini}, \citenamefont {HIll},
  \citenamefont {Jrause}, \citenamefont {Gr\:{o}blacher}, \citenamefont
  {Aspelmeyer},\ and\ \citenamefont {Painter}}]{Nature.478.89}%
  \BibitemOpen
  \bibfield  {author} {\bibinfo {author} {\bibfnamefont {J.}~\bibnamefont
  {Chan}}, \bibinfo {author} {\bibfnamefont {T.~P.~M.}\ \bibnamefont {Alegre}},
  \bibinfo {author} {\bibfnamefont {A.~H.}\ \bibnamefont {Safavi-Naeini}},
  \bibinfo {author} {\bibfnamefont {J.~T.}\ \bibnamefont {HIll}}, \bibinfo
  {author} {\bibfnamefont {A.}~\bibnamefont {Jrause}}, \bibinfo {author}
  {\bibfnamefont {S.}~\bibnamefont {Gr\:{o}blacher}}, \bibinfo {author}
  {\bibfnamefont {M.}~\bibnamefont {Aspelmeyer}},\ and\ \bibinfo {author}
  {\bibfnamefont {O.}~\bibnamefont {Painter}},\ }\href
  {https://doi.org/10.1038/nature10461} {\bibfield  {journal} {\bibinfo
  {journal} {Nature}\ }\textbf {\bibinfo {volume} {478}},\ \bibinfo {pages}
  {89} (\bibinfo {year} {2011})}\BibitemShut {NoStop}%
\bibitem [{\citenamefont {Verhagen}\ \emph {et~al.}(2012)\citenamefont
  {Verhagen}, \citenamefont {Del\'{e}glise}, \citenamefont {Weis},
  \citenamefont {Schliesser},\ and\ \citenamefont
  {Kippenberg}}]{Nature.482.63}%
  \BibitemOpen
  \bibfield  {author} {\bibinfo {author} {\bibfnamefont {E.}~\bibnamefont
  {Verhagen}}, \bibinfo {author} {\bibfnamefont {S.}~\bibnamefont
  {Del\'{e}glise}}, \bibinfo {author} {\bibfnamefont {S.}~\bibnamefont {Weis}},
  \bibinfo {author} {\bibfnamefont {A.}~\bibnamefont {Schliesser}},\ and\
  \bibinfo {author} {\bibfnamefont {T.~J.}\ \bibnamefont {Kippenberg}},\ }\href
  {https://doi.org/10.1038/nature10787} {\bibfield  {journal} {\bibinfo
  {journal} {Nature}\ }\textbf {\bibinfo {volume} {482}},\ \bibinfo {pages}
  {63} (\bibinfo {year} {2012})}\BibitemShut {NoStop}%
\bibitem [{\citenamefont {Ekinci}\ and\ \citenamefont
  {Roukes}(2005)}]{RevSciInstrum.76.061101}%
  \BibitemOpen
  \bibfield  {author} {\bibinfo {author} {\bibfnamefont {K.~L.}\ \bibnamefont
  {Ekinci}}\ and\ \bibinfo {author} {\bibfnamefont {M.~L.}\ \bibnamefont
  {Roukes}},\ }\href {https://doi.org/10.1063/1.1927327} {\bibfield  {journal}
  {\bibinfo  {journal} {Rev. Sci. Instrum.}\ }\textbf {\bibinfo {volume}
  {76}},\ \bibinfo {pages} {061101} (\bibinfo {year} {2005})}\BibitemShut
  {NoStop}%
\bibitem [{\citenamefont {Regal}\ \emph {et~al.}(2008)\citenamefont {Regal},
  \citenamefont {Teufel},\ and\ \citenamefont {Lehnert}}]{NatPhys.4.555}%
  \BibitemOpen
  \bibfield  {author} {\bibinfo {author} {\bibfnamefont {C.~A.}\ \bibnamefont
  {Regal}}, \bibinfo {author} {\bibfnamefont {J.~D.}\ \bibnamefont {Teufel}},\
  and\ \bibinfo {author} {\bibfnamefont {K.~W.}\ \bibnamefont {Lehnert}},\
  }\href {https://doi.org/10.1038/nphys974} {\bibfield  {journal} {\bibinfo
  {journal} {Nat. Phys.}\ }\textbf {\bibinfo {volume} {4}},\ \bibinfo {pages}
  {555} (\bibinfo {year} {2008})}\BibitemShut {NoStop}%
\bibitem [{\citenamefont {Teufel}\ \emph
  {et~al.}(2011{\natexlab{a}})\citenamefont {Teufel}, \citenamefont {Li},
  \citenamefont {Allman}, \citenamefont {Cicak}, \citenamefont {Sirois},
  \citenamefont {Whittaker},\ and\ \citenamefont {Simmonds}}]{Nature.471.204}%
  \BibitemOpen
  \bibfield  {author} {\bibinfo {author} {\bibfnamefont {J.~D.}\ \bibnamefont
  {Teufel}}, \bibinfo {author} {\bibfnamefont {D.}~\bibnamefont {Li}}, \bibinfo
  {author} {\bibfnamefont {M.~S.}\ \bibnamefont {Allman}}, \bibinfo {author}
  {\bibfnamefont {K.}~\bibnamefont {Cicak}}, \bibinfo {author} {\bibfnamefont
  {A.~J.}\ \bibnamefont {Sirois}}, \bibinfo {author} {\bibfnamefont {J.~D.}\
  \bibnamefont {Whittaker}},\ and\ \bibinfo {author} {\bibfnamefont {R.~W.}\
  \bibnamefont {Simmonds}},\ }\href {https://doi.org/10.1038/nature09898}
  {\bibfield  {journal} {\bibinfo  {journal} {Nature}\ }\textbf {\bibinfo
  {volume} {471}},\ \bibinfo {pages} {204} (\bibinfo {year}
  {2011}{\natexlab{a}})}\BibitemShut {NoStop}%
\bibitem [{\citenamefont {Teufel}\ \emph
  {et~al.}(2011{\natexlab{b}})\citenamefont {Teufel}, \citenamefont {Donner},
  \citenamefont {Li}, \citenamefont {Harlow}, \citenamefont {Allman},
  \citenamefont {Cicak}, \citenamefont {Sirois}, \citenamefont {Whittaker},
  \citenamefont {Lehnert},\ and\ \citenamefont {Simmonds}}]{Nature.475.359}%
  \BibitemOpen
  \bibfield  {author} {\bibinfo {author} {\bibfnamefont {J.~D.}\ \bibnamefont
  {Teufel}}, \bibinfo {author} {\bibfnamefont {T.}~\bibnamefont {Donner}},
  \bibinfo {author} {\bibfnamefont {D.}~\bibnamefont {Li}}, \bibinfo {author}
  {\bibfnamefont {J.~W.}\ \bibnamefont {Harlow}}, \bibinfo {author}
  {\bibfnamefont {M.~S.}\ \bibnamefont {Allman}}, \bibinfo {author}
  {\bibfnamefont {K.}~\bibnamefont {Cicak}}, \bibinfo {author} {\bibfnamefont
  {A.~J.}\ \bibnamefont {Sirois}}, \bibinfo {author} {\bibfnamefont {J.~D.}\
  \bibnamefont {Whittaker}}, \bibinfo {author} {\bibfnamefont {K.~W.}\
  \bibnamefont {Lehnert}},\ and\ \bibinfo {author} {\bibfnamefont {R.~W.}\
  \bibnamefont {Simmonds}},\ }\href {https://doi.org/10.1038/nature10261}
  {\bibfield  {journal} {\bibinfo  {journal} {Nature}\ }\textbf {\bibinfo
  {volume} {475}},\ \bibinfo {pages} {359} (\bibinfo {year}
  {2011}{\natexlab{b}})}\BibitemShut {NoStop}%
\bibitem [{\citenamefont {Barzanjeh}\ \emph {et~al.}(2011)\citenamefont
  {Barzanjeh}, \citenamefont {Vitali}, \citenamefont {Tombesi},\ and\
  \citenamefont {Milburn}}]{PhysRevA.84.042342}%
  \BibitemOpen
  \bibfield  {author} {\bibinfo {author} {\bibfnamefont {S.}~\bibnamefont
  {Barzanjeh}}, \bibinfo {author} {\bibfnamefont {D.}~\bibnamefont {Vitali}},
  \bibinfo {author} {\bibfnamefont {P.}~\bibnamefont {Tombesi}},\ and\ \bibinfo
  {author} {\bibfnamefont {G.~J.}\ \bibnamefont {Milburn}},\ }\href
  {https://doi.org/10.1103/PhysRevA.84.042342} {\bibfield  {journal} {\bibinfo
  {journal} {Phys. Rev. A}\ }\textbf {\bibinfo {volume} {84}},\ \bibinfo
  {pages} {042342} (\bibinfo {year} {2011})}\BibitemShut {NoStop}%
\bibitem [{\citenamefont {Barzanjeh}\ \emph {et~al.}(2012)\citenamefont
  {Barzanjeh}, \citenamefont {Abdi}, \citenamefont {Milburn}, \citenamefont
  {Tombesi},\ and\ \citenamefont {Vitali}}]{PhysRevLett.109.130503}%
  \BibitemOpen
  \bibfield  {author} {\bibinfo {author} {\bibfnamefont {S.}~\bibnamefont
  {Barzanjeh}}, \bibinfo {author} {\bibfnamefont {M.}~\bibnamefont {Abdi}},
  \bibinfo {author} {\bibfnamefont {G.~J.}\ \bibnamefont {Milburn}}, \bibinfo
  {author} {\bibfnamefont {P.}~\bibnamefont {Tombesi}},\ and\ \bibinfo {author}
  {\bibfnamefont {D.}~\bibnamefont {Vitali}},\ }\href
  {https://doi.org/10.1103/PhysRevLett.109.130503} {\bibfield  {journal}
  {\bibinfo  {journal} {Phys. Rev. Lett.}\ }\textbf {\bibinfo {volume} {109}},\
  \bibinfo {pages} {130503} (\bibinfo {year} {2012})}\BibitemShut {NoStop}%
\bibitem [{\citenamefont {Huang}(2015)}]{PhysRevA.92.043845}%
  \BibitemOpen
  \bibfield  {author} {\bibinfo {author} {\bibfnamefont {S.}~\bibnamefont
  {Huang}},\ }\href {https://doi.org/10.1103/PhysRevA.92.043845} {\bibfield
  {journal} {\bibinfo  {journal} {Phys. Rev. A}\ }\textbf {\bibinfo {volume}
  {92}},\ \bibinfo {pages} {043845} (\bibinfo {year} {2015})}\BibitemShut
  {NoStop}%
\bibitem [{\citenamefont {Andrews}\ \emph {et~al.}(2014)\citenamefont
  {Andrews}, \citenamefont {Peterson}, \citenamefont {Purdy}, \citenamefont
  {Cicak}, \citenamefont {Simmonds}, \citenamefont {Regal},\ and\ \citenamefont
  {Lehnert}}]{NatPhys.10.321}%
  \BibitemOpen
  \bibfield  {author} {\bibinfo {author} {\bibfnamefont {R.~W.}\ \bibnamefont
  {Andrews}}, \bibinfo {author} {\bibfnamefont {R.~W.}\ \bibnamefont
  {Peterson}}, \bibinfo {author} {\bibfnamefont {T.~P.}\ \bibnamefont {Purdy}},
  \bibinfo {author} {\bibfnamefont {K.}~\bibnamefont {Cicak}}, \bibinfo
  {author} {\bibfnamefont {R.~W.}\ \bibnamefont {Simmonds}}, \bibinfo {author}
  {\bibfnamefont {C.~A.}\ \bibnamefont {Regal}},\ and\ \bibinfo {author}
  {\bibfnamefont {K.~W.}\ \bibnamefont {Lehnert}},\ }\href
  {https://doi.org/10.1038/nphys2911} {\bibfield  {journal} {\bibinfo
  {journal} {Nat. Phys.}\ }\textbf {\bibinfo {volume} {10}},\ \bibinfo {pages}
  {321} (\bibinfo {year} {2014})}\BibitemShut {NoStop}%
\bibitem [{\citenamefont {Malossi}\ \emph {et~al.}(2021)\citenamefont
  {Malossi}, \citenamefont {Piergentili}, \citenamefont {Li}, \citenamefont
  {Serra}, \citenamefont {Natali}, \citenamefont {Di~Giuseppe},\ and\
  \citenamefont {Vitali}}]{PhysRevA.103.033516}%
  \BibitemOpen
  \bibfield  {author} {\bibinfo {author} {\bibfnamefont {N.}~\bibnamefont
  {Malossi}}, \bibinfo {author} {\bibfnamefont {P.}~\bibnamefont
  {Piergentili}}, \bibinfo {author} {\bibfnamefont {J.}~\bibnamefont {Li}},
  \bibinfo {author} {\bibfnamefont {E.}~\bibnamefont {Serra}}, \bibinfo
  {author} {\bibfnamefont {R.}~\bibnamefont {Natali}}, \bibinfo {author}
  {\bibfnamefont {G.}~\bibnamefont {Di~Giuseppe}},\ and\ \bibinfo {author}
  {\bibfnamefont {D.}~\bibnamefont {Vitali}},\ }\href
  {https://doi.org/10.1103/PhysRevA.103.033516} {\bibfield  {journal} {\bibinfo
   {journal} {Phys. Rev. A}\ }\textbf {\bibinfo {volume} {103}},\ \bibinfo
  {pages} {033516} (\bibinfo {year} {2021})}\BibitemShut {NoStop}%
\bibitem [{\citenamefont {Nongthombam}\ \emph {et~al.}(2021)\citenamefont
  {Nongthombam}, \citenamefont {Sahoo},\ and\ \citenamefont
  {Sarma}}]{PhysRevA.104.023509}%
  \BibitemOpen
  \bibfield  {author} {\bibinfo {author} {\bibfnamefont {R.}~\bibnamefont
  {Nongthombam}}, \bibinfo {author} {\bibfnamefont {A.}~\bibnamefont {Sahoo}},\
  and\ \bibinfo {author} {\bibfnamefont {A.~K.}\ \bibnamefont {Sarma}},\ }\href
  {https://doi.org/10.1103/PhysRevA.104.023509} {\bibfield  {journal} {\bibinfo
   {journal} {Phys. Rev. A}\ }\textbf {\bibinfo {volume} {104}},\ \bibinfo
  {pages} {023509} (\bibinfo {year} {2021})}\BibitemShut {NoStop}%
\bibitem [{\citenamefont {Martin}\ \emph {et~al.}(2004)\citenamefont {Martin},
  \citenamefont {Shnirman}, \citenamefont {Tian},\ and\ \citenamefont
  {Zoller}}]{PhysRevB.69.125339}%
  \BibitemOpen
  \bibfield  {author} {\bibinfo {author} {\bibfnamefont {I.}~\bibnamefont
  {Martin}}, \bibinfo {author} {\bibfnamefont {A.}~\bibnamefont {Shnirman}},
  \bibinfo {author} {\bibfnamefont {L.}~\bibnamefont {Tian}},\ and\ \bibinfo
  {author} {\bibfnamefont {P.}~\bibnamefont {Zoller}},\ }\href
  {https://doi.org/10.1103/PhysRevB.69.125339} {\bibfield  {journal} {\bibinfo
  {journal} {Phys. Rev. B}\ }\textbf {\bibinfo {volume} {69}},\ \bibinfo
  {pages} {125339} (\bibinfo {year} {2004})}\BibitemShut {NoStop}%
\bibitem [{\citenamefont {Vitali}\ \emph {et~al.}(2007)\citenamefont {Vitali},
  \citenamefont {Tombesi}, \citenamefont {Woolley}, \citenamefont {Doherty},\
  and\ \citenamefont {Milburn}}]{PhysRevA.76.042336}%
  \BibitemOpen
  \bibfield  {author} {\bibinfo {author} {\bibfnamefont {D.}~\bibnamefont
  {Vitali}}, \bibinfo {author} {\bibfnamefont {P.}~\bibnamefont {Tombesi}},
  \bibinfo {author} {\bibfnamefont {M.~J.}\ \bibnamefont {Woolley}}, \bibinfo
  {author} {\bibfnamefont {A.~C.}\ \bibnamefont {Doherty}},\ and\ \bibinfo
  {author} {\bibfnamefont {G.~J.}\ \bibnamefont {Milburn}},\ }\href
  {https://doi.org/10.1103/PhysRevA.76.042336} {\bibfield  {journal} {\bibinfo
  {journal} {Phys. Rev. A}\ }\textbf {\bibinfo {volume} {76}},\ \bibinfo
  {pages} {042336} (\bibinfo {year} {2007})}\BibitemShut {NoStop}%
\bibitem [{\citenamefont {Hauss}\ \emph {et~al.}(2008)\citenamefont {Hauss},
  \citenamefont {Fedorov}, \citenamefont {Andr{\'{e}}}, \citenamefont {Brosco},
  \citenamefont {Hutter}, \citenamefont {Kothari}, \citenamefont {Yeshwanth},
  \citenamefont {Shnirman},\ and\ \citenamefont
  {Sch\:{o}n}}]{NewJPhys.10.085018}%
  \BibitemOpen
  \bibfield  {author} {\bibinfo {author} {\bibfnamefont {J.}~\bibnamefont
  {Hauss}}, \bibinfo {author} {\bibfnamefont {A.}~\bibnamefont {Fedorov}},
  \bibinfo {author} {\bibfnamefont {S.}~\bibnamefont {Andr{\'{e}}}}, \bibinfo
  {author} {\bibfnamefont {V.}~\bibnamefont {Brosco}}, \bibinfo {author}
  {\bibfnamefont {C.}~\bibnamefont {Hutter}}, \bibinfo {author} {\bibfnamefont
  {R.}~\bibnamefont {Kothari}}, \bibinfo {author} {\bibfnamefont
  {S.}~\bibnamefont {Yeshwanth}}, \bibinfo {author} {\bibfnamefont
  {A.}~\bibnamefont {Shnirman}},\ and\ \bibinfo {author} {\bibfnamefont
  {G.}~\bibnamefont {Sch\:{o}n}},\ }\href
  {https://doi.org/10.1088/1367-2630/10/9/095018} {\bibfield  {journal}
  {\bibinfo  {journal} {New J. Phys.}\ }\textbf {\bibinfo {volume} {10}},\
  \bibinfo {pages} {095018} (\bibinfo {year} {2008})}\BibitemShut {NoStop}%
\bibitem [{\citenamefont {Jaehne}\ \emph {et~al.}(2008)\citenamefont {Jaehne},
  \citenamefont {Hammerer},\ and\ \citenamefont
  {Wallquist}}]{NewJPhys.10.085019}%
  \BibitemOpen
  \bibfield  {author} {\bibinfo {author} {\bibfnamefont {K.}~\bibnamefont
  {Jaehne}}, \bibinfo {author} {\bibfnamefont {K.}~\bibnamefont {Hammerer}},\
  and\ \bibinfo {author} {\bibfnamefont {M.}~\bibnamefont {Wallquist}},\ }\href
  {https://doi.org/10.1088/1367-2630/10/9/085019} {\bibfield  {journal}
  {\bibinfo  {journal} {New J. Phys.}\ }\textbf {\bibinfo {volume} {10}},\
  \bibinfo {pages} {095019} (\bibinfo {year} {2008})}\BibitemShut {NoStop}%
\bibitem [{\citenamefont {Kounalakis}\ \emph {et~al.}(2020)\citenamefont
  {Kounalakis}, \citenamefont {Blanter},\ and\ \citenamefont
  {Steele}}]{PhysRevResearch.2.023335}%
  \BibitemOpen
  \bibfield  {author} {\bibinfo {author} {\bibfnamefont {M.}~\bibnamefont
  {Kounalakis}}, \bibinfo {author} {\bibfnamefont {Y.~M.}\ \bibnamefont
  {Blanter}},\ and\ \bibinfo {author} {\bibfnamefont {G.~A.}\ \bibnamefont
  {Steele}},\ }\href {https://doi.org/10.1103/PhysRevResearch.2.023335}
  {\bibfield  {journal} {\bibinfo  {journal} {Phys. Rev. Research}\ }\textbf
  {\bibinfo {volume} {2}},\ \bibinfo {pages} {023335} (\bibinfo {year}
  {2020})}\BibitemShut {NoStop}%
\bibitem [{\citenamefont {LaHaye}\ \emph {et~al.}(2009)\citenamefont {LaHaye},
  \citenamefont {Suh}, \citenamefont {Echternach}, \citenamefont {Schwab},\
  and\ \citenamefont {Roukes}}]{Nature.459.960}%
  \BibitemOpen
  \bibfield  {author} {\bibinfo {author} {\bibfnamefont {M.~D.}\ \bibnamefont
  {LaHaye}}, \bibinfo {author} {\bibfnamefont {J.}~\bibnamefont {Suh}},
  \bibinfo {author} {\bibfnamefont {P.~M.}\ \bibnamefont {Echternach}},
  \bibinfo {author} {\bibfnamefont {K.~C.}\ \bibnamefont {Schwab}},\ and\
  \bibinfo {author} {\bibfnamefont {M.~L.}\ \bibnamefont {Roukes}},\ }\href
  {https://doi.org/10.1038/nature08093} {\bibfield  {journal} {\bibinfo
  {journal} {Nature}\ }\textbf {\bibinfo {volume} {459}},\ \bibinfo {pages}
  {960} (\bibinfo {year} {2009})}\BibitemShut {NoStop}%
\bibitem [{\citenamefont {Pirkkalainen}\ \emph {et~al.}(2013)\citenamefont
  {Pirkkalainen}, \citenamefont {Cho}, \citenamefont {Li}, \citenamefont
  {Paraoanu}, \citenamefont {Hakonen},\ and\ \citenamefont
  {Sillanp\:{a}\:{a}}}]{Nature.494.211}%
  \BibitemOpen
  \bibfield  {author} {\bibinfo {author} {\bibfnamefont {J.-M.}\ \bibnamefont
  {Pirkkalainen}}, \bibinfo {author} {\bibfnamefont {S.~U.}\ \bibnamefont
  {Cho}}, \bibinfo {author} {\bibfnamefont {J.}~\bibnamefont {Li}}, \bibinfo
  {author} {\bibfnamefont {G.~S.}\ \bibnamefont {Paraoanu}}, \bibinfo {author}
  {\bibfnamefont {P.~J.}\ \bibnamefont {Hakonen}},\ and\ \bibinfo {author}
  {\bibfnamefont {M.~A.}\ \bibnamefont {Sillanp\:{a}\:{a}}},\ }\href
  {https://doi.org/10.1038/nature11821} {\bibfield  {journal} {\bibinfo
  {journal} {Nature}\ }\textbf {\bibinfo {volume} {494}},\ \bibinfo {pages}
  {211} (\bibinfo {year} {2013})}\BibitemShut {NoStop}%
\bibitem [{\citenamefont {Chu}\ \emph {et~al.}(2017)\citenamefont {Chu},
  \citenamefont {Kharel}, \citenamefont {Renninger}, \citenamefont {Burkhart},
  \citenamefont {Frunzio}, \citenamefont {Rakich},\ and\ \citenamefont
  {Schoelkopf}}]{Science.358.199}%
  \BibitemOpen
  \bibfield  {author} {\bibinfo {author} {\bibfnamefont {Y.}~\bibnamefont
  {Chu}}, \bibinfo {author} {\bibfnamefont {P.}~\bibnamefont {Kharel}},
  \bibinfo {author} {\bibfnamefont {W.~H.}\ \bibnamefont {Renninger}}, \bibinfo
  {author} {\bibfnamefont {L.~D.}\ \bibnamefont {Burkhart}}, \bibinfo {author}
  {\bibfnamefont {L.}~\bibnamefont {Frunzio}}, \bibinfo {author} {\bibfnamefont
  {P.~T.}\ \bibnamefont {Rakich}},\ and\ \bibinfo {author} {\bibfnamefont
  {R.~J.}\ \bibnamefont {Schoelkopf}},\ }\href
  {https://doi.org/10.1126/science.aao1511} {\bibfield  {journal} {\bibinfo
  {journal} {Science}\ }\textbf {\bibinfo {volume} {358}},\ \bibinfo {pages}
  {199} (\bibinfo {year} {2017})}\BibitemShut {NoStop}%
\bibitem [{\citenamefont {Mirhosseini}\ \emph {et~al.}(2020)\citenamefont
  {Mirhosseini}, \citenamefont {Sipahigil}, \citenamefont {Kalaee},\ and\
  \citenamefont {Painter}}]{Nature.588.599}%
  \BibitemOpen
  \bibfield  {author} {\bibinfo {author} {\bibfnamefont {M.}~\bibnamefont
  {Mirhosseini}}, \bibinfo {author} {\bibfnamefont {A.}~\bibnamefont
  {Sipahigil}}, \bibinfo {author} {\bibfnamefont {M.}~\bibnamefont {Kalaee}},\
  and\ \bibinfo {author} {\bibfnamefont {O.}~\bibnamefont {Painter}},\ }\href
  {https://doi.org/10.1038/s41586-020-3038-6} {\bibfield  {journal} {\bibinfo
  {journal} {Nature}\ }\textbf {\bibinfo {volume} {588}},\ \bibinfo {pages}
  {599} (\bibinfo {year} {2020})}\BibitemShut {NoStop}%
\bibitem [{\citenamefont {Strogatz}(2004)}]{PBL.Sync.Strogatz}%
  \BibitemOpen
  \bibfield  {author} {\bibinfo {author} {\bibfnamefont {S.~H.}\ \bibnamefont
  {Strogatz}},\ }\href {https://books.google.co.in/books?id=a4bfvTMvYpcC}
  {\emph {\bibinfo {title} {Sync: The Emerging Science of Spontaneous Order}}}\
  (\bibinfo  {publisher} {Penguin Books Limited},\ \bibinfo {year}
  {2004})\BibitemShut {NoStop}%
\bibitem [{\citenamefont {Oliveira}\ and\ \citenamefont
  {Melo}(2015)}]{SciRep.5.11548}%
  \BibitemOpen
  \bibfield  {author} {\bibinfo {author} {\bibfnamefont {H.~M.}\ \bibnamefont
  {Oliveira}}\ and\ \bibinfo {author} {\bibfnamefont {L.~V.}\ \bibnamefont
  {Melo}},\ }\href {https://doi.org/10.1038/srep11548} {\bibfield  {journal}
  {\bibinfo  {journal} {Sci. Rep.}\ }\textbf {\bibinfo {volume} {5}},\ \bibinfo
  {pages} {11548} (\bibinfo {year} {2015})}\BibitemShut {NoStop}%
\bibitem [{\citenamefont {Lee}\ and\ \citenamefont
  {Sadeghpour}(2013)}]{PhysRevLett.111.234101}%
  \BibitemOpen
  \bibfield  {author} {\bibinfo {author} {\bibfnamefont {T.~E.}\ \bibnamefont
  {Lee}}\ and\ \bibinfo {author} {\bibfnamefont {H.~R.}\ \bibnamefont
  {Sadeghpour}},\ }\href {https://doi.org/10.1103/PhysRevLett.111.234101}
  {\bibfield  {journal} {\bibinfo  {journal} {Phys. Rev. Lett.}\ }\textbf
  {\bibinfo {volume} {111}},\ \bibinfo {pages} {234101} (\bibinfo {year}
  {2013})}\BibitemShut {NoStop}%
\bibitem [{\citenamefont {Hush}\ \emph {et~al.}(2015)\citenamefont {Hush},
  \citenamefont {Li}, \citenamefont {Genway}, \citenamefont {Lesanovsky},\ and\
  \citenamefont {Armour}}]{PhysRevA.91.061401}%
  \BibitemOpen
  \bibfield  {author} {\bibinfo {author} {\bibfnamefont {M.~R.}\ \bibnamefont
  {Hush}}, \bibinfo {author} {\bibfnamefont {W.}~\bibnamefont {Li}}, \bibinfo
  {author} {\bibfnamefont {S.}~\bibnamefont {Genway}}, \bibinfo {author}
  {\bibfnamefont {I.}~\bibnamefont {Lesanovsky}},\ and\ \bibinfo {author}
  {\bibfnamefont {A.~D.}\ \bibnamefont {Armour}},\ }\href
  {https://doi.org/10.1103/PhysRevA.91.061401} {\bibfield  {journal} {\bibinfo
  {journal} {Phys. Rev. A}\ }\textbf {\bibinfo {volume} {91}},\ \bibinfo
  {pages} {061401} (\bibinfo {year} {2015})}\BibitemShut {NoStop}%
\bibitem [{\citenamefont {Walter}\ \emph {et~al.}(2015)\citenamefont {Walter},
  \citenamefont {Nunnenkamp},\ and\ \citenamefont
  {Bruder}}]{AnnPhysBerl.527.131}%
  \BibitemOpen
  \bibfield  {author} {\bibinfo {author} {\bibfnamefont {S.}~\bibnamefont
  {Walter}}, \bibinfo {author} {\bibfnamefont {A.}~\bibnamefont {Nunnenkamp}},\
  and\ \bibinfo {author} {\bibfnamefont {C.}~\bibnamefont {Bruder}},\ }\href
  {https://doi.org/10.1002/andp.201400144} {\bibfield  {journal} {\bibinfo
  {journal} {Ann. Phys.}\ }\textbf {\bibinfo {volume} {527}},\ \bibinfo {pages}
  {131} (\bibinfo {year} {2015})}\BibitemShut {NoStop}%
\bibitem [{\citenamefont {L\"{o}rch}\ \emph {et~al.}(2016)\citenamefont
  {L\"{o}rch}, \citenamefont {Amitai}, \citenamefont {Nunnenkamp},\ and\
  \citenamefont {Bruder}}]{PhysRevLett.117.073601}%
  \BibitemOpen
  \bibfield  {author} {\bibinfo {author} {\bibfnamefont {N.}~\bibnamefont
  {L\"{o}rch}}, \bibinfo {author} {\bibfnamefont {E.}~\bibnamefont {Amitai}},
  \bibinfo {author} {\bibfnamefont {A.}~\bibnamefont {Nunnenkamp}},\ and\
  \bibinfo {author} {\bibfnamefont {C.}~\bibnamefont {Bruder}},\ }\href
  {https://doi.org/10.1103/PhysRevLett.117.073601} {\bibfield  {journal}
  {\bibinfo  {journal} {Phys. Rev. Lett.}\ }\textbf {\bibinfo {volume} {117}},\
  \bibinfo {pages} {073601} (\bibinfo {year} {2016})}\BibitemShut {NoStop}%
\bibitem [{\citenamefont {Li}\ \emph {et~al.}(2017)\citenamefont {Li},
  \citenamefont {Zhang}, \citenamefont {Li},\ and\ \citenamefont
  {Song}}]{CommunNonlinearSciNumerSimulat.42.121}%
  \BibitemOpen
  \bibfield  {author} {\bibinfo {author} {\bibfnamefont {W.}~\bibnamefont
  {Li}}, \bibinfo {author} {\bibfnamefont {F.}~\bibnamefont {Zhang}}, \bibinfo
  {author} {\bibfnamefont {C.}~\bibnamefont {Li}},\ and\ \bibinfo {author}
  {\bibfnamefont {H.}~\bibnamefont {Song}},\ }\href
  {https://doi.org/10.1016/j.cnsns.2016.05.015} {\bibfield  {journal} {\bibinfo
   {journal} {Commun. Nonlinear Sci. Numer. Simulat.}\ }\textbf {\bibinfo
  {volume} {42}},\ \bibinfo {pages} {121} (\bibinfo {year} {2017})}\BibitemShut
  {NoStop}%
\bibitem [{\citenamefont {Koppenh\"{o}fer}\ and\ \citenamefont
  {Roulet}(2019)}]{PhysRevA.99.043804}%
  \BibitemOpen
  \bibfield  {author} {\bibinfo {author} {\bibfnamefont {M.}~\bibnamefont
  {Koppenh\"{o}fer}}\ and\ \bibinfo {author} {\bibfnamefont {A.}~\bibnamefont
  {Roulet}},\ }\href {https://doi.org/10.1103/PhysRevA.99.043804} {\bibfield
  {journal} {\bibinfo  {journal} {Phys. Rev. A}\ }\textbf {\bibinfo {volume}
  {99}},\ \bibinfo {pages} {043804} (\bibinfo {year} {2019})}\BibitemShut
  {NoStop}%
\bibitem [{\citenamefont {Koppenh\"ofer}\ \emph {et~al.}(2020)\citenamefont
  {Koppenh\"ofer}, \citenamefont {Bruder},\ and\ \citenamefont
  {Roulet}}]{PhysRevResearch.2.023026}%
  \BibitemOpen
  \bibfield  {author} {\bibinfo {author} {\bibfnamefont {M.}~\bibnamefont
  {Koppenh\"ofer}}, \bibinfo {author} {\bibfnamefont {C.}~\bibnamefont
  {Bruder}},\ and\ \bibinfo {author} {\bibfnamefont {A.}~\bibnamefont
  {Roulet}},\ }\href {https://doi.org/10.1103/PhysRevResearch.2.023026}
  {\bibfield  {journal} {\bibinfo  {journal} {Phys. Rev. Research}\ }\textbf
  {\bibinfo {volume} {2}},\ \bibinfo {pages} {023026} (\bibinfo {year}
  {2020})}\BibitemShut {NoStop}%
\bibitem [{\citenamefont {Zhirov}\ and\ \citenamefont
  {Shepelyansky}(2008)}]{PhysRevLett.100.014101}%
  \BibitemOpen
  \bibfield  {author} {\bibinfo {author} {\bibfnamefont {O.~V.}\ \bibnamefont
  {Zhirov}}\ and\ \bibinfo {author} {\bibfnamefont {D.~L.}\ \bibnamefont
  {Shepelyansky}},\ }\href {https://doi.org/10.1103/PhysRevLett.100.014101}
  {\bibfield  {journal} {\bibinfo  {journal} {Phys. Rev. Lett.}\ }\textbf
  {\bibinfo {volume} {100}},\ \bibinfo {pages} {014101} (\bibinfo {year}
  {2008})}\BibitemShut {NoStop}%
\bibitem [{\citenamefont {Amitai}\ \emph {et~al.}(2017)\citenamefont {Amitai},
  \citenamefont {L\"{o}rch}, \citenamefont {Nunnenkamp}, \citenamefont
  {Walter},\ and\ \citenamefont {Bruder}}]{PhysRevA.95.053858}%
  \BibitemOpen
  \bibfield  {author} {\bibinfo {author} {\bibfnamefont {E.}~\bibnamefont
  {Amitai}}, \bibinfo {author} {\bibfnamefont {N.}~\bibnamefont {L\"{o}rch}},
  \bibinfo {author} {\bibfnamefont {A.}~\bibnamefont {Nunnenkamp}}, \bibinfo
  {author} {\bibfnamefont {S.}~\bibnamefont {Walter}},\ and\ \bibinfo {author}
  {\bibfnamefont {C.}~\bibnamefont {Bruder}},\ }\href
  {https://doi.org/10.1103/PhysRevA.95.053858} {\bibfield  {journal} {\bibinfo
  {journal} {Phys. Rev. A}\ }\textbf {\bibinfo {volume} {95}},\ \bibinfo
  {pages} {053858} (\bibinfo {year} {2017})}\BibitemShut {NoStop}%
\bibitem [{\citenamefont {Bowen}\ and\ \citenamefont
  {Milburn}(2015)}]{TaylorFrancis.QuantumOptomechanics.Bowen}%
  \BibitemOpen
  \bibfield  {author} {\bibinfo {author} {\bibfnamefont {W.~P.}\ \bibnamefont
  {Bowen}}\ and\ \bibinfo {author} {\bibfnamefont {G.~J.}\ \bibnamefont
  {Milburn}},\ }\href {https://books.google.co.in/books?id=xqlcrgEACAAJ} {\emph
  {\bibinfo {title} {Quantum Optomechanics}}}\ (\bibinfo  {publisher} {Taylor
  \& Francis},\ \bibinfo {year} {2015})\BibitemShut {NoStop}%
\bibitem [{\citenamefont {Brun}(2002)}]{AmJPhys.70.719}%
  \BibitemOpen
  \bibfield  {author} {\bibinfo {author} {\bibfnamefont {T.~A.}\ \bibnamefont
  {Brun}},\ }\href {https://doi.org/10.1119/1.1475328} {\bibfield  {journal}
  {\bibinfo  {journal} {American Journal of Physics}\ }\textbf {\bibinfo
  {volume} {70}},\ \bibinfo {pages} {719} (\bibinfo {year} {2002})}\BibitemShut
  {NoStop}%
\bibitem [{\citenamefont {Johansson}\ \emph {et~al.}(2012)\citenamefont
  {Johansson}, \citenamefont {Nation},\ and\ \citenamefont
  {Nori}}]{ComputPhysCommun.183.1760}%
  \BibitemOpen
  \bibfield  {author} {\bibinfo {author} {\bibfnamefont {J.~R.}\ \bibnamefont
  {Johansson}}, \bibinfo {author} {\bibfnamefont {P.~D.}\ \bibnamefont
  {Nation}},\ and\ \bibinfo {author} {\bibfnamefont {F.}~\bibnamefont {Nori}},\
  }\href {https://doi.org/10.1016/j.cpc.2012.02.021} {\bibfield  {journal}
  {\bibinfo  {journal} {Comput. Phys. Commun}\ }\textbf {\bibinfo {volume}
  {183}},\ \bibinfo {pages} {1760} (\bibinfo {year} {2012})}\BibitemShut
  {NoStop}%
\bibitem [{\citenamefont {Johansson}\ \emph {et~al.}(2013)\citenamefont
  {Johansson}, \citenamefont {Nation},\ and\ \citenamefont
  {Nori}}]{ComputPhysCommun.184.1234}%
  \BibitemOpen
  \bibfield  {author} {\bibinfo {author} {\bibfnamefont {J.~R.}\ \bibnamefont
  {Johansson}}, \bibinfo {author} {\bibfnamefont {P.~D.}\ \bibnamefont
  {Nation}},\ and\ \bibinfo {author} {\bibfnamefont {F.}~\bibnamefont {Nori}},\
  }\href {https://doi.org/10.1016/j.cpc.2012.11.019} {\bibfield  {journal}
  {\bibinfo  {journal} {Comput. Phys. Commun}\ }\textbf {\bibinfo {volume}
  {184}},\ \bibinfo {pages} {1234} (\bibinfo {year} {2013})}\BibitemShut
  {NoStop}%
\end{thebibliography}%

\end{document}